\documentstyle[12pt]{article}
\input{tcilatex}
\begin{document}

\begin{titlepage}
\begin{center}
{\hbox to\hsize{
\hfill \bf hep-th/0307243 }}
{\hbox to\hsize{\hfill July 2003 }}

\bigskip
\vspace{3\baselineskip}

{\Large \bf

Deformed N=1 supersymmetry\\}

\bigskip

\bigskip

{\bf Masud Chaichian$^{\mathrm{a}}$ and  Archil Kobakhidze$^{\mathrm{a,b}}$ \\}
\smallskip

{ \small \it
$^{\mathrm{a}}$High Energy Physics Division, Department of Physical Sciences, 
University of Helsinki $\&$ \\
Helsinki Institute of Physics, FIN-00014 Helsinki, Finland\\ 
$^{\mathrm{b}}$Andronikashvili Institute of Physics, Georgian Academy of Sciences, 
GE-380077 Tbilisi, Georgia\\}

\bigskip

\vspace*{.5cm}

{\bf Abstract}\\
\end{center}
\noindent
We consider a deformation of N=1 four dimensional Minkowski superspace  
where odd coordinates $\theta^{\alpha}$ do not anticommute. 
We define supersymmetric and associative star 
product and show how the remaining (anti)commutation 
relations among the superspace coordinates are modified. In particular, the even coordinates  
do not commute as well. We also study chiral and vector superfields and their 
interactions. Suprisingly we find  that ordinary undeformed N=1 supersymmetric 
field theories are compatible with the deformed supersymmetry considered.
\bigskip

\bigskip

\end{titlepage}

\section{Introduction}

It is widely believed that further progress in our understanding of
elementary particles and fundamental interactions is ultimately related with
deeper understanding of the actual structure of space-time at short
distances. The idea that usual four space-time coordinates could be
supplemented by anticommuting spinorial coordinates $\theta ^{\alpha }$, $%
\overline{\theta }^{\stackrel{.}{\alpha }}$, 
\begin{eqnarray}
\left\{ \theta ^{\alpha },\theta ^{\beta }\right\} &=&\left\{ \overline{%
\theta }^{\stackrel{.}{\alpha }},\overline{\theta }^{\stackrel{.}{\beta }%
}\right\} =0,  \label{1} \\
\left\{ \theta ^{\alpha },\overline{\theta }^{\stackrel{.}{\beta }}\right\}
&=&0,  \label{2}
\end{eqnarray}
leads to a notion of (N=1) superspace. Field theories on such extension of
space-time have improved ultra-violet (UV) behavior compared to the usual
fields theories due to the remarkable symmetry between the fields with Bose
and Fermi statistics, called supersymmetry (SUSY). Largely because of this
N=1 SUSY is currently considered as a prime candidate for the fundamental
particles and interactions beyond the celebrated Standard Model.

Another interesting theoretical approach, which originally was also thought
to be responsible for the curing of UV divergences, is space-time
noncommutativity. In the past few years noncommutative field theories (see
e.g. \cite{1} for a review) have attracted enormous interest after the
realization that the noncommutativity arises naturally as a low energy limit
of string theory in the background of constant antisymmetric NS-NS B-field 
\cite{2}. However, noncommutative field theories show up some pathological
features which prevent to apply them to particle physics (see e.g. \cite{3}
and \cite{4} for the attempts to construct noncommutative Standard Model).
The most difficult problem is the occurrence of infrared divergences \cite{5}
which perhaps indicate that pure field theoretical description is
incomplete. Other difficulties are related to peculiar properties of
noncommutative gauge groups and their representations which are reflected in
charge quantization problem \cite{6} and to anomalies \cite{7} which make
construction of desired chiral theories problematic.

Straightforward SUSY extension of noncommutative field theories where the
odd superspace coordinates satisfy the usual (anti)commutation relations 
\cite{8} is of little help in this situation. However, it has been also
realized that SUSY is actually compatible with more general
(anti)commutation relations \cite{9,10,11}. More recently, non-trivial
(anti)commutation relations for the odd coordinates has been obtained from
the string theory in the constant graviphoton and/or gravitino backgrounds 
\cite{12,13,14}. This observation initiates subsequently recent studies \cite
{14,15,16,17} of field theories with deformed SUSY mainly in Euclidean space.%
\footnote{%
After the completion of this work recent paper \cite{18} appeared which
considers the general case of deformed Poisson brackets including N=1
supersymmetric case.}

Since our main interest is possible implications of deformed SUSY in
particle physics, in the present paper we study a particular deformation of
N=1 four dimensional Minkowski superspace. In particular, instead of
ordinary anticommutation relations (\ref{1}) we assume that odd superspace
coordinates do not anticommute. In subsequent sections we define
supersymmetric and associative Weyl-Moyal-type star products and show how
the consistency dictates modification of (anti)commutation relations of
other superspace coordinates. In particular, the even coordinates do not
commute as well. Then we define the relevant superfields and construct
invariant actions. Somewhat surprisingly we will find that ordinary
undeformed supersymmetric field theories are compatible with the deformed
N=1 supersymmetry considered.

\section{Deformed superspace}

Consider N=1 four dimensional Minkowski superspace with a set of coordinates 
$\left( x^{\mu },\theta ^{\alpha },\overline{\theta }^{\stackrel{.}{\alpha }%
}\right) $. We would like to consider deformation of anticommutation
relations (\ref{1}). Namely, assume 
\begin{equation}
\left\{ \widehat{\theta }^{\alpha },\widehat{\theta }^{\beta }\right\}
=C^{\alpha \beta },  \label{3}
\end{equation}
where $C^{\alpha \beta }\left( =C^{\beta \alpha }\right) $ are complex
constants. Since in Minkowski space-time $\left( \widehat{\theta }^{\alpha
}\right) ^{\dagger }=\widehat{\overline{\theta }}^{\stackrel{.}{\alpha }}$,
we also have 
\begin{equation}
\left\{ \widehat{\overline{\theta }}^{\stackrel{.}{\alpha }},\widehat{%
\overline{\theta }}^{\stackrel{.}{\beta }}\right\} =\overline{C}^{\stackrel{.%
}{\alpha }\stackrel{.}{\beta }},  \label{4}
\end{equation}
$\overline{C}^{\stackrel{.}{\alpha }\stackrel{.}{\beta }}=\left( C^{\beta
\alpha }\right) ^{\dagger }$. Because of (\ref{3}), product of functions of $%
\widehat{\theta }^{\alpha }$ should be correspondingly ordered. As in the
ordinary nonsommutative case, this can be done by a suitably defined
Weyl-Moyal-type star product of functions of ordinary anticommuting $\theta $%
's. We define (see also \cite{18}): 
\begin{eqnarray}
f\left( \widehat{\theta }\right) g\left( \widehat{\theta }\right) &\equiv
&f\left( \theta \right) \star g\left( \theta \right) \stackrel{\text{def}}{=}%
f\left( \theta \right) \exp \left[ -\overleftarrow{D}_{\alpha }\frac{%
C^{\alpha \beta }}{2}\overrightarrow{D}_{\beta }\right] g\left( \theta
\right)  \nonumber \\
&=&f\left( \theta \right) \left[ 1-\overleftarrow{D}_{\alpha }\frac{%
C^{\alpha \beta }}{2}\overrightarrow{D}_{\beta }-\frac{1}{16}\det \overline{C%
}\overleftarrow{D}^{2}\overrightarrow{D}^{2}\right] g\left( \theta \right) ,
\label{5}
\end{eqnarray}
where $\overrightarrow{D}_{\alpha }$ and $\overleftarrow{D}_{\alpha }$ are
left and right covariant derivatives, respectively, 
\begin{equation}
\overrightarrow{D}_{\alpha }=\frac{\overrightarrow{\partial }}{\partial
\theta ^{\alpha }}+i\sigma _{\alpha \stackrel{.}{\alpha }}^{\mu }\overline{%
\theta }^{\stackrel{.}{\alpha }}\frac{\overrightarrow{\partial }}{\partial
x^{\mu }},\hspace{1cm}\overleftarrow{D}_{\alpha }=-\frac{\overleftarrow{%
\partial }}{\partial \theta ^{\alpha }}+i\sigma _{\alpha \stackrel{.}{\alpha 
}}^{\mu }\overline{\theta }^{\stackrel{.}{\alpha }}\frac{\overleftarrow{%
\partial }}{\partial x^{\mu }}  \label{6}
\end{equation}
and $\overrightarrow{D}^{2}=2\overrightarrow{D}_{2}\overrightarrow{D}_{1}$, $%
\overleftarrow{D}^{2}=2\overleftarrow{D}_{1}\overleftarrow{D}_{2}$ .%
\footnote{%
We follow the conventions of Wess and Bagger \cite{19}.} Then we have $%
\left\{ \theta ^{\alpha },\theta ^{\beta }\right\} _{\star }=C^{\alpha \beta
}$. Here and below the subscript $\star $ ($\overline{\star }$) means that $%
\star $-product ($\overline{\star }$-product) is involved.{\bf \ }Taking
Hermitian conjugate of (\ref{5}) we define conjugate star product of
functions of $\overline{\theta }^{\stackrel{.}{\alpha }}$: 
\begin{eqnarray}
g\left( \widehat{\overline{\theta }}\right) f\left( \widehat{\overline{%
\theta }}\right) &\equiv &\overline{g}\left( \overline{\theta }\right) 
\overline{\star }\overline{f}\left( \overline{\theta }\right) \stackrel{%
\text{def}}{=}\overline{g}\left( \overline{\theta }\right) \exp \left[ -%
\overleftarrow{\overline{D}}_{\stackrel{.}{\alpha }}\frac{\overline{C}^{%
\stackrel{.}{\alpha }\stackrel{.}{\beta }}}{2}\overrightarrow{\overline{D}}_{%
\stackrel{.}{\beta }}\right] \overline{f}\left( \overline{\theta }\right) 
\nonumber \\
&=&\overline{g}\left( \overline{\theta }\right) \left[ 1-\overleftarrow{%
\overline{D}}_{\stackrel{.}{\alpha }}\frac{\overline{C}^{\stackrel{.}{\alpha 
}\stackrel{.}{\beta }}}{2}\overrightarrow{\overline{D}}_{\stackrel{.}{\beta }%
}-\frac{1}{16}\det \overline{C}\overleftarrow{\overline{D}}^{2}%
\overrightarrow{\overline{D}}^{2}\right] \overline{f}\left( \overline{\theta 
}\right) ,  \label{7}
\end{eqnarray}
where $\overrightarrow{\overline{D}}_{\stackrel{.}{\alpha }}=\left( 
\overleftarrow{D}_{\alpha }\right) ^{\dagger }$, $\overleftarrow{\overline{D}%
}_{\stackrel{.}{\alpha }}=\left( \overrightarrow{D}_{\alpha }\right)
^{\dagger }$: 
\begin{equation}
\overrightarrow{\overline{D}}_{\stackrel{.}{\alpha }}=-\frac{\overrightarrow{%
\partial }}{\partial \overline{\theta }^{\stackrel{.}{\alpha }}}-i\theta
^{\alpha }\sigma _{\alpha \stackrel{.}{\alpha }}^{\mu }\frac{\overrightarrow{%
\partial }}{\partial x^{\mu }},\hspace{1cm}\overleftarrow{\overline{D}}%
_{\alpha }=\frac{\overleftarrow{\partial }}{\partial \overline{\theta }^{%
\stackrel{.}{\alpha }}}-i\theta ^{\alpha }\sigma _{\alpha \stackrel{.}{%
\alpha }}^{\mu }\frac{\overleftarrow{\partial }}{\partial x^{\mu }}.
\label{8}
\end{equation}
So we have, $\left\{ \overline{\theta }^{\stackrel{.}{\alpha }},\overline{%
\theta }^{\stackrel{.}{\beta }}\right\} _{\overline{\star }}=\overline{C}^{%
\stackrel{.}{\alpha }\stackrel{.}{\beta }}$. Note, that the product $f\left( 
\widehat{\theta }\right) g\left( \widehat{\overline{\theta }}\right) $ is
not deformed neither under the $\star $-product nor under the $\overline{%
\star }$-product, 
\begin{equation}
f\left( \widehat{\theta }\right) g\left( \widehat{\overline{\theta }}\right)
=f\left( \theta \right) g\left( \overline{\theta }\right) .  \label{9}
\end{equation}

Obviously, the covariant derivatives satisfy the following aticommutation
relations 
\begin{eqnarray}
\left\{ \overrightarrow{D}_{\alpha },\overrightarrow{D}_{\beta }\right\}
_{\star } &=&0,\hspace{0.5cm}\left\{ \overrightarrow{\overline{D}}_{%
\stackrel{.}{\alpha }},\overrightarrow{\overline{D}}_{\stackrel{.}{\beta }%
}\right\} _{\overline{\star }}=0,  \label{a1} \\
\left\{ \overrightarrow{D}_{\alpha },\overrightarrow{\overline{D}}_{%
\stackrel{.}{\beta }}\right\} _{\star ,\overline{\star }} &=&-2i\sigma
_{\alpha \stackrel{.}{\beta }}^{\mu }\frac{\partial }{\partial x^{\mu }}.
\label{a2}
\end{eqnarray}
However, 
\begin{equation}
\left\{ \overrightarrow{D}_{\alpha },\overrightarrow{D}_{\beta }\right\} _{%
\overline{\star }}=\sigma _{\alpha \stackrel{.}{\alpha }}^{\mu }\sigma
_{\beta \stackrel{.}{\beta }}^{\nu }\overline{C}^{\stackrel{.}{\alpha }%
\stackrel{.}{\beta }}\frac{\partial }{\partial x^{\mu }}\frac{\partial }{%
\partial x^{\nu }},\hspace{0.5cm}\left\{ \overrightarrow{\overline{D}}_{%
\stackrel{.}{\alpha }},\overrightarrow{\overline{D}}_{\stackrel{.}{\beta }%
}\right\} _{\star }=C^{\alpha \beta }\sigma _{\alpha \stackrel{.}{\alpha }%
}^{\mu }\sigma _{\beta \stackrel{.}{\beta }}^{\nu }\frac{\partial }{\partial
x^{\mu }}\frac{\partial }{\partial x^{\nu }}.  \label{a3}
\end{equation}
This means that $D_{\alpha }(\overline{D}_{\stackrel{.}{\alpha }})$ does not
act as a derivation with respect to $\overline{\star }(\star )$ -product,
but it does with respect to $\star (\overline{\star })$ -product. Then it
follows that the subalgebra of chiral (antichiral) superfields is not closed
with respect to $\star (\overline{\star })$ -product.

The star products (\ref{5}) and (\ref{7}) are associative and supersymmetric 
\cite{18}. The associativity can be directly verified using (\ref{5}) and (%
\ref{7}) and it actually is the result of anticommutation relations (\ref{a1}%
). Also one can easily see that the supercharges 
\begin{eqnarray}
\overrightarrow{Q}_{\alpha } &=&\overrightarrow{D}_{\alpha }-2i\sigma
_{\alpha \stackrel{.}{\alpha }}^{\mu }\overline{\theta }^{\stackrel{.}{%
\alpha }}\frac{\overrightarrow{\partial }}{\partial x^{\mu }}  \label{a4} \\
\overrightarrow{\overline{Q}}_{\stackrel{.}{\alpha }} &=&\overrightarrow{%
\overline{D}}_{\stackrel{.}{\alpha }}+2i\theta ^{\alpha }\sigma _{\alpha 
\stackrel{.}{\alpha }}^{\mu }\frac{\overrightarrow{\partial }}{\partial
x^{\mu }},  \label{a5}
\end{eqnarray}
act as a derivations with respect to both $\star $ and $\overline{\star }$
products as a result of anticommutation relations: 
\begin{equation}
\left\{ \overrightarrow{Q}_{\alpha },\overrightarrow{D}_{\beta }\right\}
_{\star }=\left\{ \overrightarrow{\overline{Q}}_{\stackrel{.}{\alpha }},%
\overrightarrow{D}_{\beta }\right\} _{\star }=\left\{ \overrightarrow{Q}%
_{\alpha },\overrightarrow{\overline{D}}_{\stackrel{.}{\beta }}\right\} _{%
\overline{\star }}=\left\{ \overrightarrow{\overline{Q}}_{\stackrel{.}{%
\alpha }},\overrightarrow{\overline{D}}_{\stackrel{.}{\beta }}\right\} _{%
\overline{\star }}=0.  \label{a6}
\end{equation}

Having defined the above star products, the remaining (anti)commutation
relations among the superspace coordinates follow immediately:\bigskip 
\begin{eqnarray}
\star \text{ -- deformation} &:&\hspace{2cm} 
\begin{array}{ll}
\left\{ \theta ^{\alpha },\overline{\theta }^{\stackrel{.}{\beta }}\right\}
_{\star }=0, & \left\{ \overline{\theta }^{\stackrel{.}{\alpha }},\overline{%
\theta }^{\stackrel{.}{\beta }}\right\} _{\star }=0,
\end{array}
\nonumber \\
&&  \nonumber \\
&&\hspace{2cm} 
\begin{array}{ll}
\left[ x^{\mu },\theta ^{\alpha }\right] _{\star }=iC^{\alpha \beta }\sigma
_{\beta \stackrel{.}{\beta }}^{\mu }\overline{\theta }^{\stackrel{.}{\beta }%
}, & \left[ x^{\mu },\overline{\theta }^{\stackrel{.}{\alpha }}\right]
_{\star }=0,
\end{array}
\label{10} \\
&&  \nonumber \\
&&\hspace{2cm} 
\begin{array}{ll}
\left[ x^{\mu },x^{\nu }\right] _{\star }=C^{\alpha \beta }\sigma _{\alpha
}^{\mu \nu \text{ }\gamma }\epsilon _{\gamma \beta }\overline{\theta }%
\overline{\theta } & 
\end{array}
\nonumber
\end{eqnarray}
\bigskip

\begin{eqnarray}
\overline{\star }\text{ -- deformation} &:&\text{{}}\hspace{2cm} 
\begin{array}{ll}
\left\{ \theta ^{\alpha },\overline{\theta }^{\stackrel{.}{\beta }}\right\}
_{\overline{\star }}=0, & \left\{ \theta ^{\alpha },\theta ^{\beta }\right\} 
\overline{_{\star }}=0,
\end{array}
\nonumber \\
&&  \nonumber \\
&&\hspace{2cm} 
\begin{array}{ll}
\left[ x^{\mu },\theta ^{\alpha }\right] _{\overline{\star }}=0, & \left[
x^{\mu },\overline{\theta }^{\stackrel{.}{\alpha }}\right] _{\overline{\star 
}}=i\theta ^{\beta }\sigma _{\beta \stackrel{.}{\beta }}^{\mu }\overline{C}^{%
\stackrel{.}{\beta }\stackrel{.}{\alpha }},
\end{array}
\label{11} \\
&&  \nonumber \\
&&\hspace{2cm} 
\begin{array}{ll}
\left[ x^{\mu },x^{\nu }\right] _{\overline{\star }}=\theta \theta \epsilon
_{\stackrel{.}{\alpha }\stackrel{.}{\gamma }}\overline{\sigma }_{\stackrel{.%
}{\text{ }\beta }}^{\mu \nu \stackrel{.}{\gamma }}\overline{C}^{\stackrel{.}{%
\alpha }\stackrel{.}{\beta }}. & 
\end{array}
\nonumber
\end{eqnarray}
In particular one sees that even coordinates $x^{\mu }$ do not commute as
well.

For a product of generic superfields we can adopt either $\star $-product or 
$\overline{\star }$-product. Peculiar property of the above star products is
that their are not Hermitian, i.e. 
\begin{equation}
\left( f\left( x,\theta ,\overline{\theta }\right) \star g\left( x,\theta ,%
\overline{\theta }\right) \right) ^{\dagger }=\overline{g}\left( x,\theta ,%
\overline{\theta }\right) \overline{\star }\overline{f}\left( x,\theta ,%
\overline{\theta }\right) .  \label{12}
\end{equation}
However, the hermiticity of Lagrangians ensures that both star deformations
are physically equivalent. In what follows in the rest of the paper we adopt 
$\star $-deformation for definiteness.

\section{Chiral superfields}

The chiral (antichiral) superfield is defined to satisfy $\overline{D}_{%
\stackrel{.}{\alpha }}\Phi =0$ $\left( D_{\alpha }\overline{\Phi }=0\right) $%
. In the case of undeformed supersymmetry chiral (antichiral) superfields
form a close subalgebra, i.e. any product of chiral (antichiral) superfields
is a chiral (antichiral) superfield. However, if we consider $\star $%
-deformed multiplication this is no longer true. Namely, as it can be easily
checked using (\ref{5}), chiral superfields do not form the closed
subalgebra under the $\star $-product, while antichiral superfields do. {\it %
Vice versa}, a $\overline{\star }$-product of antichiral superfields is not
an antichiral superfield, while $\overline{\star }$-product of chiral
superfields is a chiral superfield. This means that we can write only
antichiral superpotential of the form: 
\begin{equation}
\overline{{\cal W}}_{(\star )}=\sum_{n}g_{n}\overline{\Phi }_{\star
}^{n}=g_{2}\overline{\Phi }\star \overline{\Phi }+g_{3}\overline{\Phi }\star 
\overline{\Phi }\star \overline{\Phi }+...  \label{15}
\end{equation}
and its Hermitian conjugated chiral superpotential of the form: 
\begin{equation}
{\cal W}_{(\overline{\star })}=\sum_{n}\overline{g}_{n}\Phi _{\overline{%
\star }}^{n}=\overline{g}_{2}\Phi \overline{\star }\Phi +\overline{g}%
_{3}\Phi \overline{\star }\Phi \overline{\star }\Phi +...  \label{16}
\end{equation}
Moreover, $\star $-product ($\overline{\star }$-product) of antichiral
(chiral) superfields is equivalent to the ordinary product, that is to say $%
\overline{{\cal W}}_{(\star )}({\cal W}_{(\overline{\star })})=\overline{%
{\cal W}}({\cal W})$, and thus the superpotential of the undeformed
Wess-Zumino model is consistent with deformed supersymmetry under
consideration. Also it is easy to see, the $\star $-product of chiral and
antichiral superfields (the kinetic term of the Wess-Zumino model) is not
deformed 
\begin{equation}
\Phi \star \overline{\Phi }=\Phi \left[ 1-\overleftarrow{D}_{\alpha }\frac{%
C^{\alpha \beta }}{2}\overrightarrow{D}_{\beta }-\frac{1}{16}\det C%
\overleftarrow{D}^{2}\overrightarrow{D}^{2}\right] \overline{\Phi }=\Phi 
\overline{\Phi },  \label{17}
\end{equation}
because of $D_{\alpha }\overline{\Phi }=0$. Thus we conclude that the
ordinary undeformed Wess-Zumino model is fully compatible with deformed
supersymmetry!\footnote{%
In \cite{18}, it has been pointed out that the deformed N=1 supersymmetry
does allow to write D-terms of the $\star $-product of chiral superfields, $%
\left. \overline{D}^{2}D^{2}\sum_{n}g_{n}\Phi _{\star }^{n}\right| _{\theta =%
\overline{\theta }=0}$(along with their Hermitian conjugate terms), which
give a deformation of the Wess-Zumino model involving derivatives of the
component fields. Obviously, such kind of terms in the classical
(undeformed) limit, $C\rightarrow 0$, vanish.}

\section{Vector superfields}

Consider now SUSY gauge theory with arbitrary gauge group. The element of
the SUSY gauge group is given by a chiral superfield $\exp \left( i\Lambda
\right) $, where $\Lambda =\Lambda ^{a}T^{a}$ are the chiral superfields ($%
\overline{D}_{\stackrel{.}{\alpha }}\Lambda ^{a}=0$) and $T^{a}$ are the
generators of the group. The antichiral element of the SUSY gauge group then
is $\exp \left( -i\overline{\Lambda }\right) ,$ ($D_{\alpha }\overline{%
\Lambda }^{a}=0$). Once again, in order to preserve chirality of those
superfields we should consider the following deformations: $\exp _{\overline{%
\star }}\left( i\Lambda \right) $ and $\exp _{\star }\left( -i\overline{%
\Lambda }\right) $, which are equivalent to the corresponding undeformed
superfields. This means that contrary to the case of ordinary noncommutative
gauge theories the algebra of gauge group is not deformed and we have no
restrictions on possible gauge groups and their representations. Actually,
we will see momentarily that like the Wess-Zumino model discussed in the
previous section, the ordinary undeformed gauge theories are also compatible
with deformed SUSY.

The gauge fields are residing in a vector superfield $V$, which is a
Hermitian matrix, $V^{\dagger }=V$. The gauge symmetry acts 
\begin{eqnarray}
e_{\star }^{V} &\rightarrow &\exp _{\star }\left( -i\overline{\Lambda }%
\right) \star e_{\star }^{V}\overline{\star }\exp _{\overline{\star }}\left(
i\Lambda \right) =  \nonumber \\
&&\exp \left( -i\overline{\Lambda }\right) e_{\star }^{V}\exp \left(
i\Lambda \right) ,  \label{18}
\end{eqnarray}
i.e. as in the case of undeformed SUSY. This means that we can use ordinary
Wess-Zumino gauge, where 
\begin{equation}
V(x,\theta ,\overline{\theta })=-\sigma ^{\mu }\overline{\theta }A_{\mu
}(x)+i\theta \theta \overline{\theta }\overline{\lambda }(x)-i\overline{%
\theta }\overline{\theta }\theta \lambda (x)+\frac{1}{2}\theta \theta 
\overline{\theta }\overline{\theta }D(x).  \label{19}
\end{equation}
Correspondingly we define chiral and antichiral field strength superfields: 
\begin{eqnarray}
W_{\alpha }^{(\overline{\star })} &=&-\frac{1}{4}\overline{D}\overline{D}e_{%
\overline{\star }}^{-V}\overline{\star }D_{\alpha }e_{\overline{\star }}^{V},
\nonumber \\
\overline{W}_{\stackrel{.}{\alpha }}^{(\star )} &=&-\frac{1}{4}DDe_{\star
}^{V}\star \overline{D}_{\stackrel{.}{\alpha }}e_{\star }^{-V}.  \label{20}
\end{eqnarray}
Using (\ref{18}), one sees that they transform under the gauge
transformations as in the usual case 
\begin{eqnarray}
W_{\alpha }^{(\overline{\star })} &\rightarrow &\exp \left( -i\Lambda
\right) W_{\alpha }^{(\overline{\star })}\exp \left( i\Lambda \right) 
\nonumber \\
\overline{W}_{\stackrel{.}{\alpha }}^{(\star )} &\rightarrow &\exp \left( -i%
\overline{\Lambda }\right) \overline{W}_{\stackrel{.}{\alpha }}^{(\star
)}\exp \left( i\overline{\Lambda }\right) .  \label{21}
\end{eqnarray}
In fact, it is easy to see that field strength superfields (\ref{20}) are
not deformed. Indeed, 
\begin{eqnarray}
W_{\alpha }^{(\overline{\star })} &=&-\frac{1}{4}\overline{D}\overline{D}e_{%
\overline{\star }}^{-V}\overline{\star }D_{\alpha }e_{\overline{\star }%
}^{V},=-\frac{1}{4}\overline{D}\overline{D}\left( D_{\alpha }V-\frac{1}{2}%
\left[ D_{\alpha }V,V\right] _{\overline{\star }}\right) =-\frac{1}{4}%
\overline{D}\overline{D}\left( D_{\alpha }V-\frac{1}{2}\left[ D_{\alpha
}V,V\right] \right) -  \nonumber \\
&&\frac{1}{8}\overline{D}\overline{D}\overrightarrow{\overline{D}}_{%
\stackrel{.}{\beta }}\left( \frac{\overline{C}^{\stackrel{.}{\beta }%
\stackrel{.}{\gamma }}}{2}\left[ D_{\alpha }{}{}V\overrightarrow{\overline{D}%
}_{\stackrel{.}{\gamma }}V+V\overrightarrow{\overline{D}}_{\stackrel{.}{%
\gamma }}D_{\alpha }{}{}V\right] +\frac{\det \overline{C}}{16}\left[ 
\overrightarrow{\overline{D}}^{\gamma }D_{\alpha }V\overrightarrow{\overline{%
D}}^{2}V-\overrightarrow{\overline{D}}^{\stackrel{.}{\gamma }}V%
\overrightarrow{\overline{D}}^{2}D_{\alpha }V\right] \right)  \nonumber \\
&=&W_{\alpha }  \label{22}
\end{eqnarray}
and similarly for $\overline{W}_{\stackrel{.}{\alpha }}^{(\star )},$ $%
\overline{W}_{\stackrel{.}{\alpha }}^{(\star )}=\overline{W}_{\stackrel{.}{%
\alpha }}.$ Then the invariant Lagrangian 
\begin{equation}
{\cal L}_{\star SYM}=\int d^{2}\theta \frac{1}{4}W\overline{\star }W+\int
d^{2}\overline{\theta }\frac{1}{4}\overline{W}\star \overline{W}  \label{23}
\end{equation}
is indeed nothing but the Lagrangian of ordinary super-Yang-Mills theory!

\section{Conclusions}

We have discussed the deformation of N=1 four dimensional Minkowski
superspace assuming that the odd superspace coordinates $\theta ^{\alpha }$
do not anticommute. As it has been recently shown in \cite{14}, in Euclidean
space this deformation can be described by a star product which respects
only N=$\frac{1}{2}$ supersymmetry. In Minkowski space-time this deformation
can be described by N=1 supersymmetric $\star $-product which is
non-Hermitian. Hermiticity of Lagrangians ensures that this $\star $%
-deformation and its conjugated $\overline{\star }$-deformation are
physically equivalent. The (anti)commutation relations among other
superspace coordinates are also modified (see, (\ref{10}, \ref{11})). In
particular, ordinary space-time coordinates do not commute as well.

We have defined chiral and vector superfields and discussed their
interactions. Although in general one is able to write corrections to the
classical (undeformed) N=1 Wess-Zumino model, but it is amusing that the
ordinary Wess-Zumino model alone is fully compatible with deformed
supersymmetry. We have demonstrated that the algebra of gauge groups are not
deformed and hence the ordinary super-Yang-Mills theories are also
consistent with the deformation of supersymmetry we have considered.

\subparagraph{Acknowledgments.}

This work was supported by the Academy of Finland under the Project No.
54023.

\end{document}